\documentstyle[12pt]{article}
\newcommand{\text}{\rm}

\begin{document}

\title{{\bf BRST Cohomology of }$N=2${\bf \ Super-Yang-Mills Theory in }$4D$}
\author{{\bf A. Tanzini}$^{a,b}${\bf , O. S. Ventura}$^{a,b}${\bf ,} \and {\bf %
L.C.Q.Vilar}$^b${\bf , S.P. Sorella}$^b${\bf \ } \and \vspace{1.5mm} \\
$^a$CBPF, Centro Brasileiro de Pesquisas F\'{\i}sicas \\
Departamento de Campos e Part\'{\i}culas\\
Rua Xavier Sigaud 150, 22290-180 Urca \\
Rio de Janeiro, Brazil.\vspace{1.5mm}\\
$^b$UERJ, Universidade do Estado do Rio de Janeiro\\
Departamento de F\'{\i}sica Te\'{o}rica, Instituto de F\'{\i}sica\\
Rua S\~ao Francisco Xavier, 524\\
20550-013, Maracan\~{a}, Rio de Janeiro, Brazil. \and UERJ/DFT - 06 - 2000%
\vspace{2.0cm}}
\maketitle

\begin{abstract}
The BRST\ cohomology of the $N=2$ supersymmetric Yang-Mills theory in four
dimensions is discussed by making use of the twisted version of the $N=2$
algebra. By the introduction of a set of suitable constant ghosts associated
to the generators of $N=2$, the quantization of the model can be done by
taking into account both gauge invariance and supersymmetry. In particular,
we show how the twisted $N=2$ algebra can be used to obtain in a
straightforward way the relevant cohomology classes. Moreover, we shall be
able to establish a very useful relationship between the local gauge
invariant polynomial $tr\phi ^2$ and the complete $N=2$ Yang-Mills action.
This important relation can be considered as the first step towards a fully
algebraic proof of the one-loop exactness of the $N=2$ $\beta $-function.

\setcounter{page}{0}\thispagestyle{empty}
\end{abstract}





\vfill\newpage\ \makeatother

\renewcommand{\theequation}{\thesection.\arabic{equation}}

\section{\ Introduction\-}

The $N=2$ super-Yang-Mills field theory has a long history and many
interesting properties, both at the perturbative and nonperturbative
level, as for instance the one-loop exactness of its $\beta $-function.

After Witten's work {\it \cite{tym}, } $N=2$
super-Yang-Mills became known to be deeply related to the topological
Yang-Mills theory (TYM). In fact, as pointed out by {\it \cite{tym}, } 
the TYM action 
can be seen  as the twisted version of the euclidean $N=2$ supersymmetric
Yang-Mills theory {\it \cite{mar}. }The twisted supersymmetric charges of $%
N=2$ then lead to a scalar, a vector and a self-dual tensor charge, which
are the symmetry generators of the twisted action. The final relationship of
the twisted $N=2$ theory with TYM is done by identifying the ${\cal %
R}$-charge of the twisted fields and generators with the ghost number. This
means that the matter fields of $N=2$ acquire the status of ghost fields and
the nilpotent scalar operator obtained from the twist of the supersymmetric 
generators is interpreted as a BRST-like operator. The effect of this
identification results in the independence of the theory from any scale, as
the BRST cohomology becomes completely trivial. In fact, it was shown that
TYM can be fully obtained from the gauge fixing of a surface term (the
Pontryagin index) {\it \cite{lp,bs}. }Furthermore, the Witten observables of
the TYM\ can be recovered through the formalism of the equivariant
cohomology {\it \cite{stora1,stora2}.}

More recently, an important progress in the algebraic quantization of
supersymmetric field theories in the Wess-Zumino gauge has been achieved.
This goal has been accomplished by collecting the gauge symmetry and all the
supersymmetry generators into an extended BRST operator, as shown by {\it 
\cite{white} }in the{\it \ }case of{\it \ }$N=4${\it \ }Yang-Mills.
Moreover, the case of $N=2$ super-Yang-Mills has also been successfully
worked out, with the result that its BRST cohomology is in fact nontrivial,
allowing for a nonvanishing beta function {\it \cite{magg}}. However, a
fully algebraic proof of the nonrenormalization properties of the $N=2$ $%
\beta $-function is still lacking.

Our attitude in this paper will be to take into further analysis the twisted
version of the conventional $N=2$ supersymmetric euclidean Yang-Mills
theory, without identifying the ${\cal R}$-charge with the ghost number. The
twisting mechanism has the advantage of clarifying the topological structure
which underlies the $N=2$ super-Yang-Mills field theory. This will allow us
to analyse the BRST structure of the $N=2$ super-Yang-Mills within the
framework of the descent equations, providing a better understanding of the
finiteness properties displayed by the model.

At this stage, we will refer to this version as the twisted theory (TSYM)
and leave the denomination of TYM for the cohomological theory obtained
after the aforementioned identification. It is worth mentioning here that the 
renormalization of Witten's topological Yang-Mills has been discussed by the authors 
{\it \cite{horne, al} }, who have been able to show that the topological character 
of TYM is preserved at the quantum level.  

Notice that the classical action for the $N=2$ twisted theory will be just 
Witten's TYM action, but now with all
fields interpreted as gauge or matter fields. In this way, the quantization
of the twisted theory can proceed along the same line of the quantization of
the supersymmetric models in the Wess-Zumino gauge {\it \cite{white,magg}}.
As we shall see, this procedure will require to take into account the full
set of symmetries (gauge and supersymmetry). The contact with the
cohomological formulations of TYM {\it \cite{lp,bs} }will then be
established. It is worth underling that the requirement of analyticity in
the constant ghosts will play a fundamental role in order to obtain
nontrivial cohomology classes. In fact, we shall be able to show that one
can move from a nontrivial theory to the cohomological TYM by giving up of
this analyticity condition. In this perturbative approach, we will have the
possibility to show that the unique invariant counterterm of the $N=2$
theory can be fully characterized by analysing the solutions of the set of
descent equations corresponding to the integrated BRST\ cohomology. As a
consequence, we shall be able to show that the full $N=2$ Yang-Mills action
turns out to be related to the invariant polynomial $tr\phi ^2$. This will
allow us to guess on the origin of the one-loop exactness of the $\beta 
$-function in $N=2$ supersymmetric Yang-Mills theory {\it \cite{beta}}.

The paper is organized as follows. In the next section we present a simple
review of the twisting mechanism, making the connection of TYM with $N=2$
super-Yang-Mills. In Section three we proceed with the quantization of the
twisted theory. Section four is devoted to the renormalization of the model.
Finally, we shall make contact with the results already existing in the
literature and we shall draw a possible path toward the algebraic proof of
the nonrenormalization theorem for the $N=2$ $\beta $-function.

\section{The Twisted Action}

{\it \ }Following {\it \cite{tym}, } the TSYM classical action is given by 
\begin{equation}
\begin{array}{l}
{\cal S}_{TSYM}=\displaystyle\frac 1{g^2}tr\displaystyle\int d^4x\;\left(
\frac 12F_{\mu \nu }^{+}F^{+\mu \nu }\;+\frac 12\overline{\phi }\left\{ \psi
^\mu ,\psi _\mu \right\} \right. \\ 
-\chi ^{\mu \nu }(D_\mu \psi _\nu -D_\nu \psi _\mu )^{+}+\eta D_\mu \psi
^\mu \;-\displaystyle\frac 12\overline{\phi }D_\mu D^\mu \phi \; \\ 
\displaystyle\left. -\frac 12\phi \left\{ \chi ^{\mu \nu },\chi _{\mu \nu
}\right\} \;-\frac 18\left[ \phi ,\eta \right] \eta -\frac 1{32}\left[ \phi ,%
\overline{\phi }\right] \left[ \phi ,\overline{\phi }\right] \right) \;,
\end{array}
\label{tym}
\end{equation}
where $g$ is the {\it unique} coupling constant and $F_{\mu \nu }^{+}$ is
the self-dual part of the Yang-Mills field strength. The three fields \ $%
\left( \chi _{\mu \nu },\psi _\mu ,\eta \right) $ in the expression $\left( 
\ref{tym}\right) $ are anticommuting, with $\chi _{\mu \nu }$ self-dual, and 
$\left( \phi ,\overline{\phi }\right) $ are commuting complex scalar fields, 
$\overline{\phi }$ being assumed to be the complex conjugate of $\phi $. Of
course, TSYM being a gauge theory, is left invariant by the gauge
transformations

\begin{eqnarray}
\delta _\epsilon ^gA_\mu &=&-D_\mu \epsilon \;,  \label{g-transf} \\
\delta _\epsilon ^g\lambda &=&\left[ \epsilon ,\lambda \right]
\;,\;\;\;\;\lambda =\chi ,\psi ,\eta ,\phi ,\overline{\phi }\;.  \nonumber
\end{eqnarray}

It is easily checked that the kinetic terms in the action{\it \ }$\left( \ref
{tym}\right) $ {\it \ }corresponding to the fields{\it \ }$\left( \chi ,\psi
,\eta ,\phi ,\overline{\phi }\right) $ are nondegenerate, so that these
fields have well defined propagators. The only degeneracy is that related to
the pure Yang-Mills term{\it \ }$F_{\mu \nu }^{+}F^{+\mu \nu }$. Therefore,
from{\it \ }eq{\it .}$\left( \ref{g-transf}\right) $ one is led to interpret 
{\it \ }$\left( \chi ,\psi ,\eta ,\phi ,\overline{\phi }\right) $ as
ordinary matter fields, in spite of the unconventional tensorial character of%
{\it \ }$\left( \chi _{\mu \nu },\psi _\mu \right) $. We assign to $(A,\chi
,\psi ,\eta ,\phi ,\overline{\phi })$ the dimensions $(1,3/2,3/2,3/2,1,1)$
and ${\cal R}$-charges $(0,-1,1,-1,2,-2)$, so that the TSYM action $\left( 
\ref{tym}\right) $ has vanishing total ${\cal R}$-charge. Let us emphasize
once more that we do not identify the ${\cal R}$-charge with the ghost
number, so that we avoid the cohomological interpretation which leads to TYM
theory. The action $\left( \ref{tym}\right) $ has to be regarded just as the
twisted version of $N=2$ super-Yang-Mills theory.

For a better understanding of this point, let us briefly review the twisting
procedure of the $N=2$ supersymmetric algebra in flat euclidean space-time 
{\it \cite{mar}}. In the absence of central extension, the $N=2$
supersymmetry in the Wess-Zumino gauge is characterized by $8$ charges $%
\left( {\cal Q}_{\;\alpha }^i,\overline{{\cal Q}}_{\;\dot{\alpha}}^j\right) $
obeying the following relations 
\begin{equation}
\begin{array}{l}
\left\{ {\cal Q}_{\;\alpha }^i,\overline{{\cal Q}}_{j\dot{\alpha}}\right\}
=\delta _j^i\partial _{\alpha \dot{\alpha}}+{\rm gauge\;transf.}+{\rm eqs.}\;%
{\rm mot.}\;, \\ 
\left\{ {\cal Q}_{\;\alpha }^i,{\cal Q}_{\;\beta }^j\right\} =\left\{ 
\overline{{\cal Q}}_{\dot{\alpha}}^i,\overline{{\cal Q}}_{\dot{\beta}%
}^j\right\} ={\rm gauge\;transf.}+{\rm eqs.\;mot.}\;.
\end{array}
\label{wz-n=2}
\end{equation}
where $(\alpha ,\dot{\alpha})=1,2$ are the spinor indices, $(i,j)=1,2\;$the
internal $SU(2)$ indices labelling the different charges of $N=2,$ and $%
\partial _{\alpha \dot{\alpha}}=\left( \sigma ^\mu \right) _{\alpha \dot{%
\alpha}}\partial _\mu $,$\;\sigma ^\mu \;$being the Pauli matrices. The
special feature of $N=2$ is that both spinor and internal indices run from $%
1\;$to $2$, making it possible to identify the index $i$ with one of the two
spinor indices $(\alpha ,\dot{\alpha}).$ This corresponds to substitute the $%
SU\left( 2\right) _L$ factor of the Lorentz simmetry group of the theory
with the diagonal subgroup $SU(2)_A^{\prime }={\rm diag}\left( SU\left(
2\right) _L\times SU\left( 2\right) _A\right) $. It is precisely this
identification which defines the twisting procedure {\it \cite{mar}}.
Identifying therefore the internal index $i\;$with the spinor index $\alpha $%
, we can construct now the following twisted generators\ $\left( \delta
,\delta _\mu ,\delta _{\mu \nu }\right) ,$

\begin{eqnarray}
\delta &=&\frac 1{\sqrt{2}}\varepsilon ^{\alpha \beta }{\cal Q}_{\beta
\alpha }\;,\;\delta _\mu =\frac 1{\sqrt{2}}\overline{{\cal Q}}_{\alpha \dot{%
\alpha}}\;\left( \overline{\sigma }_\mu \right) ^{\dot{\alpha}\alpha }\;,
\label{twist-g} \\
\delta _{\mu \nu } &=&\frac 1{\sqrt{2}}(\sigma _{\mu \nu })^{\alpha \beta }%
{\cal Q}_{\beta \alpha }=-\delta _{\nu \mu \;}.  \nonumber
\end{eqnarray}
Notice that the generators $\delta _{\mu \nu }$ are self-dual due to the
fact that the matrices $\sigma _{\mu \nu }$ are self-dual in euclidean
space-time. In terms of these generators, the $N=2$ susy algebra reads now 
\begin{equation}
\begin{array}{c}
\delta ^2={\rm gauge}\;{\rm transf.\;}+\;{\rm eqs.}\;{\rm of\;motion}\;, \\ 
\left\{ \delta ,\delta _\mu \right\} =\partial _\mu \;+{\rm gauge\;transf.\;}%
+\;{\rm eqs.\;of\;motion}\;, \\ 
\left\{ \delta _\mu ,\delta _\nu \right\} ={\rm gauge\;transf.\;}+\;{\rm %
eqs.\;of\;motion}\;,
\end{array}
\label{wz-talg-1}
\end{equation}
\begin{equation}
\begin{array}{c}
\left\{ \delta ,\delta _{\mu \nu }\right\} =\left\{ \delta _{\mu \nu
},\delta _{\rho \sigma }\right\} ={\rm gauge\;transf.\;}+\;{\rm %
eqs.\;of\;motion}\;, \\ 
\left\{ \delta _\mu ,\delta _{\rho \sigma }\right\} =-\varepsilon _{\mu \rho
\sigma \nu }\partial ^\nu -g_{\mu [\rho }\partial _{\sigma ]}+{\rm %
gauge\;transf.\;}+\;{\rm eqs.\;of\;motion}\;.
\end{array}
\label{wz-talg-2}
\end{equation}
The algebraic structure realized by the generators $(\delta ,\delta _\mu )$
in eq.$\left( \ref{wz-talg-1}\right) $ is typical of the topological models 
{\it \cite{tft,brt}}. In this case the vector charge $\delta _\mu $, usually
called vector supersymmetry, is known to play an important role in the
derivation of the ultraviolet finiteness properties of the topological
models and in the construction of their observables {\it \cite{tft}}.

Let us now turn to the relationship between Witten's TYM and the $N=2$
Yang-Mills theory, and show, in particular, that TYM has the same field
content of the $N=2$ Yang-Mills theory in the Wess-Zumino gauge. The minimal 
$N=2$ supersymmetric\ pure Yang-Mills theory is described by a multiplet\
which, in the Wess-Zumino gauge, contains {\it \cite{magg,mar}}: a gauge
field $A_\mu $, two spinors $\psi _\alpha ^i\;i=1,2$, their conjugate $%
\overline{\psi }_{\dot{\alpha}}^i$, and two scalars $\phi ,\overline{\phi }%
\;(\overline{\phi }$ being the complex conjugate of $\phi )$. All these
fields are in the adjoint representation of the gauge group. We proceed by
applying the previous twisting procedure to the $N=2\;$Wess-Zumino vector
multiplet $(A_\mu ,\psi _\alpha ^i,\overline{\psi }_{\dot{\alpha}}^i,\phi ,%
\overline{\phi })$.\ Identifying then the internal index $i$ with the spinor
index $\alpha $, it is very easy to see that the spinor $\overline{\psi }_{%
\dot{\alpha}}^i\;$can be related to an anticommuting vector $\psi _\mu ,$\
i.e

\begin{equation}
\overline{\psi }_{\dot{\alpha}}^i\stackrel{twist}{\longrightarrow }\overline{%
\psi }_{\alpha \dot{\alpha}}\longrightarrow \psi _\mu =(\overline{\sigma }%
_\mu )^{\dot{\alpha}\alpha }\overline{\psi }_{\alpha \dot{\alpha}}\;.
\label{t-spinor}
\end{equation}
Concerning now the fields $\psi _\beta ^i\;$we have$\;\psi _\beta ^i%
\stackrel{twist}{\longrightarrow }\psi _{\alpha \beta }=\psi _{(\alpha \beta
)}+\psi _{[\alpha \beta ]}\;,\;\psi _{(\alpha \beta )}\;$and $\psi _{[\alpha
\beta ]}\;$being respectively symmetric and antisymmetric in the spinor
indices $\alpha ,\beta $. To $\psi _{[\alpha \beta ]}\;$we associate an
anticommuting scalar field $\eta ,$ while $\psi _{(\alpha \beta )}\;$turns
out to be related to an antisymmetric self-dual field $\chi _{\mu \nu }\;$%
through 
\begin{equation}
\begin{array}{c}
\psi _{[\alpha \beta ]}\longrightarrow \eta =\varepsilon ^{\alpha \beta
}\psi _{[\alpha \beta ]}\;, \\ 
\psi _{(\alpha \beta )}\longrightarrow \chi _{\mu \nu }=\widetilde{\chi }%
_{\mu \nu }=(\sigma _{\mu \nu })^{\alpha \beta }\psi _{(\alpha \beta )}\;.
\end{array}
\label{eta-field}
\end{equation}
Therefore, the twisting procedure allows to replace the $N=2\;$Wess-Zumino
multiplet $(A_\mu ,\psi _\alpha ^i,\overline{\psi }_{\dot{\alpha}}^i,\phi ,%
\overline{\phi })$ by the twisted multiplet $(A_\mu ,\psi _\mu ,\chi _{\mu
\nu },\eta ,\phi ,\overline{\phi })$ whose field content is precisely that
of the TSYM\ action $\left( \ref{tym}\right) $. The same holds for the $N=2$%
\ pure Yang-Mills action {\it \cite{mar}}, 
\[
{\cal S}_{YM}^{N=2}(A_\mu ,\psi _\alpha ^i,\overline{\psi }_{\dot{\alpha}%
}^i,\phi ,\overline{\phi })\stackrel{twist}{\longrightarrow }{\cal S}%
_{TSYM}(A_\mu ,\psi _\mu ,\chi _{\mu \nu },\eta ,\phi ,\overline{\phi }). 
\]
Thus the TSYM\ comes in fact from the twisted version of the ordinary $N=2$
Yang-Mills in the Wess-Zumino gauge. This important point, already
underlined by Witten in his original work {\it \cite{tym}}, deserves a few
clarifying remarks in order to make contact with the results on topological
field theories obtained in the recent years.

The first observation is naturally related to the existence of further
symmetries of the TSYM\ action $\left( \ref{tym}\right) $. According to the
previous analysis, we conclude that the TSYM will be left invariant by the
twisted generators $\left( \delta ,\delta _\mu ,\delta _{\mu \nu }\right) $.
In fact, it is easy to check that the twisted scalar generator $\delta $ \
corresponds to Witten's fermionic symmetry $\delta _{{\cal W}}$\ {\it \cite
{tym}} 
\begin{equation}
\begin{array}{ccc}
\delta _{{\cal W}}A_\mu =\psi _\mu \;, & \delta _{{\cal W}}\psi _\mu =-D_\mu
\phi \;, & \delta _{{\cal W}}\phi =0\;, \\ 
\delta _{{\cal W}}\chi _{\mu \nu }=F_{\mu \nu }^{+}\;, & \delta _{{\cal W}}%
\overline{\phi }=2\eta \;, & \delta _{{\cal W}}\eta =\frac 12\left[ \phi ,%
\overline{\phi }\right] \;.
\end{array}
\label{d-transf}
\end{equation}
The action of the vector generator on the set of fields is given by

\begin{eqnarray}
\delta _\mu A_\nu &=&\frac 12\chi _{\mu \nu }+\frac 18g_{\mu \nu }\eta \;,
\label{vsusy} \\
\delta _\mu \psi _\nu &=&F_{\mu \nu }-\frac 12F_{\mu \nu }^{+}-\frac
1{16}g_{\mu \nu }[\phi ,\overline{\phi }]\;,  \nonumber \\
\delta _\mu \eta &=&\frac 12D_\mu \overline{\phi }\;,  \nonumber \\
\delta _\mu \chi _{\sigma \tau } &=&\frac 18(\varepsilon _{\mu \sigma \tau
\nu }D^\nu \overline{\phi }+g_{\mu \sigma }D_\tau \overline{\phi }-g_{\mu
\tau }D_\sigma \overline{\phi })\;,  \nonumber \\
\delta _\mu \phi &=&-\psi _\mu \;,  \nonumber \\
\delta _\mu \overline{\phi } &=&0\;,  \nonumber
\end{eqnarray}
and 
\begin{equation}
\delta _{{\cal W}}{\cal S}_{TSYM}=\delta _\mu {\cal S}_{TSYM}=0\;.
\label{v-susy-inv}
\end{equation}
We underline here that the form of the TSYM action $\left( \ref{tym}\right) $
is not completely specified by the fermionic symmetry$\;\delta _{{\cal W}}$.
In other words,{\it \ }$\left( \ref{tym}\right) $ is not the most general
gauge invariant action compatible with the{\it \ }$\delta _{{\cal W}}$%
-invariance.{\it \ }Nevertheless, it turns out to be uniquely characterized
by $\delta _\mu $. The conditions $\left( \ref{v-susy-inv}\right) $ fix all
the relative numerical coefficients of the Witten's action $\left( \ref{tym}%
\right) ,$ allowing, in particular, for a single coupling constant. This
feature will be recovered in the renormalizability analysis of the model.
The last generator, $\delta _{\mu \nu }$, will reproduce, together with the
operators $\delta _{{\cal W}},\delta _\mu $, the complete $N=2\;$susy
algebra $\left( \ref{wz-talg-1}\right) ,\left( \ref{wz-talg-2}\right) $. The
reasons why we do not actually take in further account the transformations $%
\delta _{\mu \nu }$ are due partly to the fact that, as previously remarked,
the TSYM action is already uniquely fixed by the$\;(\delta _{{\cal W}}$, $%
\delta _\mu )-$symmetries and partly to the fact that the generator $\delta
_{\mu \nu }\;$turns out to be trivially realized on the fields in terms of
the $\delta _{{\cal W}}$-transformations {\it \cite{vitoria}}.

The second remark is related to the standard perturbative Feynman diagram
computations. From the equivalence between ${\cal S}_{YM}^{N=2}$ and ${\cal S%
}_{TSYM}$ it is very tempting to argue that the values of quantities like
the $\beta $-function should be the same when computed in the ordinary $N=2$%
\ Yang-Mills and in the twisted version. After all, at least at the
perturbative level, the twisting procedure has the effect of a linear change
of variables on the fields. The computation of the\ one loop $\beta $%
-function for the twisted theory has indeed been performed by R. Brooks et
al. {\it \cite{beta}}. As expected, the result agrees with that of the
untwisted $N=2\;$Yang-Mills.

\section{\ Quantizing the Twisted Theory} 

Before facing the problem of quantizing the theory, let us spend a few words on the 
strategy which will be adopted in the following. As we have already seen in the previous section, the twisted algebra of the generators of $N=2$ closes on the translations only 
on shell and up to gauge transformations, due to the use of the Wess-Zumino gauge. We are dealing therefore with an open algebra, whose quantization requires the introduction of an appropriate set of antifields. Instead of making use of the convential Batalin-Fradkin-Vilkovisky (BFV) approach, we shall proceed here with a slight different procedure, already successfully used in the case of $N=4$ super Yang-Mills  {\it \cite{white}}, and completely equivalent to the BFV method. 

We shall first begin by looking at
an extended BRST operator ${\cal Q}$ which turns out to be nilpotent on
shell and which will define the gauge-fixed action. The construction of the final Slavnov-Taylor equation (or master equation) will be thus achieved by adding to the gauge-fixed  action 
a suitable set of antifields, including in particular terms which are quadratic in the antifields, as required by the BFV procedure in the case of open algebras. 
 
In order to obtain the BRST operator ${\cal Q}$ we introduce the Faddeev-Popov ghost field $c\;$%
corresponding to the gauge transformations $\left( \ref{g-transf}\right) $, 
\begin{equation}
\begin{array}{ccc}
sA_\mu =-D_\mu c\;, & sc=c^2\;, & s\phi =\left[ c,\phi \right] \;, \\ 
s\psi _\mu =\left\{ c,\psi _\mu \right\} \;, & s\chi _{\mu \nu }=\left\{
c,\chi _{\mu \nu }\right\} \;, & s\eta =\left\{ c,\eta \right\} \;, \\ 
s\overline{\phi }=\left[ c,\overline{\phi }\right] \;, &  & 
\end{array}
\label{s-op=transf}
\end{equation}

\begin{equation}
s{\cal S}_{TSYM}=0\,,\;\;s^2=0\;.  \label{s-op-inv}
\end{equation}
We associate now to each generator entering the algebra $\left( \ref
{wz-talg-1}\right) $, namely $\delta _{{\cal W}}$, $\delta _\mu \;$and $%
\partial _\mu $, the\ constant ghost parameters $\left( \omega ,\varepsilon
^\mu ,v^\mu \right) $ respectively, defining, in this way, the extended BRST
operator

\begin{equation}
{\cal Q=}s+\omega \delta _{{\cal W}}+\varepsilon ^\mu \delta _\mu +v^\mu
\partial _\mu -\omega \varepsilon ^\mu \frac \partial {\partial v^\mu }\;.
\label{Q-op}
\end{equation}

Now, we need to define the action of the four generators $s,\delta _{{\cal W}%
},\delta _\mu $ and $\partial _\mu $\ on the ghosts $\left( c,\omega
,\varepsilon ^\mu ,v^\mu \right) $. Let us work out in detail the case of
the two operators $s$\ and $\delta _{{\cal W}}$. Working out eq.$\left( \ref
{wz-talg-1}\right) $ explicitly, one looks then for an operator $(s+\omega
\delta _{{\cal W}})$ nilpotent on ${\rm \;}(A_\mu ,\psi _\mu ,\eta ,\phi ,%
\overline{\phi },c,\omega )$ and nilpotent on shell on $\chi _{\mu \nu }$ 
{\it \cite{vitoria}}. After a little experiment, it is not difficult to
convince oneself that these conditions are indeed verified by defining the
action of $s$\ and $\delta _{{\cal W}}\;$on the ghost $(c,\omega )\;$as

\begin{equation}
s\omega =0\;,\;\;\;\delta _{{\cal W}}\omega =0\;,\;\;\;\delta _{{\cal W}%
}c=-\omega \phi \;.  \label{s-om-d-def}
\end{equation}

The above procedure can be now easily repeated in order to include in the
game also the operators $\delta _\mu $ and $\partial _\mu $. The final
result is that the extension of the operator ${\cal Q}$ on the ghosts $%
\left( c,\omega ,\varepsilon ^\mu ,v^\mu \right) $ is found to be

\begin{eqnarray}
{\cal Q}c &=&c^2-\omega ^2\phi -\omega \varepsilon ^\mu A_\mu +\frac{%
\varepsilon ^2}{16}\overline{\phi }+v^\mu \partial _\mu c\;,  \label{Q-ext}
\\
{\cal Q}\omega &=&0\;,\;\;\;{\cal Q}\varepsilon ^\mu =0\;,\;\;\;{\cal Q}%
v^\mu =-\omega \varepsilon ^\mu \;,  \nonumber
\end{eqnarray}
with

\begin{equation}
{\cal Q}^2=0\;\;\;\;{\rm on\;\;\;\;}\left( A,\phi ,\overline{\phi },\eta
,c,\omega ,\varepsilon ,v \right) \;,  \label{QQ-1}
\end{equation}
and

\begin{eqnarray}
{\cal Q}^2\psi _\sigma &=&\frac{g^2}4\omega \varepsilon ^\mu \frac{\delta 
{\cal S}_{TYM}}{\delta \chi ^{\mu \sigma }}  \label{QQ-2} \\
&&+\frac{g^2}{32}\varepsilon ^\mu \varepsilon ^\nu \left( g_{\mu \sigma }%
\frac{\delta {\cal S}_{TYM}}{\delta \psi ^\nu }+g_{\nu \sigma }\frac{\delta 
{\cal S}_{TYM}}{\delta \psi ^\mu }-2g_{\mu \nu }\frac{\delta {\cal S}_{TYM}}{%
\delta \psi ^\sigma }\right) \;,  \nonumber
\end{eqnarray}

\begin{eqnarray}
{\cal Q}^2\chi _{\sigma \tau } &=&-\frac{g^2}2\omega ^2\frac{\delta {\cal S}%
_{TYM}}{\delta \chi ^{\sigma \tau }}  \label{QQ-3} \\
&&+\frac{g^2}8\omega \varepsilon ^\mu \left( \varepsilon _{\mu \sigma \tau
\nu }\frac{\delta {\cal S}_{TYM}}{\delta \psi _\nu }+g_{\mu \sigma }\frac{%
\delta {\cal S}_{TYM}}{\delta \psi ^\tau }-g_{\mu \tau }\frac{\delta {\cal S}%
_{TYM}}{\delta \psi ^\sigma }\right) \;.  \nonumber
\end{eqnarray}

For the usefulness of the reader, we give in the tables \ I and II the
quantum numbers and the Grassmanian characters of all the fields and
constant ghosts. We observe that the grading is chosen to be the sum of the
ghost number and of the ${\cal R}$-charge.

\[
\begin{array}{c}
\begin{tabular}{|c|c|c|c|c|c|c|}
\hline
& $A_\mu $ & $\chi _{\mu \nu }$ & $\psi _\mu $ & $\eta $ & $\phi $ & $%
\overline{\phi }$ \\ \hline
dim. & $1$ & $3/2$ & $3/2$ & $3/2$ & $1$ & $1$ \\ \hline
${\cal R}$-charge & $0$ & $-1$ & $1$ & $-1$ & $2$ & $-2$ \\ \hline
gh-number & $0$ & $0$ & $0$ & $0$ & $0$ & $0$ \\ \hline
$\ $nature & $comm.$ & $ant.$ & $ant.$ & $ant.$ & $comm.$ & $comm.$ \\ \hline
\end{tabular}
\\ 
{\rm {Table 1: Quantum numbers}}
\end{array}
\]

\[
\begin{array}{c}
\begin{tabular}{|c|c|c|c|c|}
\hline
& $c$ & $\omega $ & $\varepsilon _\mu $ & $v_\mu $ \\ \hline
$dim.$ & $0$ & $-1/2$ & $-1/2$ & $-1$ \\ \hline
${\cal R}$-charge & $0$ & $-1$ & $1$ & $0$ \\ \hline
gh-number & $1$ & $1$ & $1$ & $1$ \\ \hline
nature & $ant.$ & $comm.$ & $comm.$ & $ant.$ \\ \hline
\end{tabular}
\\ 
{\rm {Table 2: Quantum numbers}}
\end{array}
\]

The construction of the gauge fixing term is now straightforward. We
introduce an antighost $\overline{c}$ and a Lagrangian multiplier $b$
transforming as {\it \cite{white,magg}}$\;$%
\begin{equation}
{\cal Q}\overline{c}\;=b+v^\mu \partial _\mu \overline{c}\;,\;\;{\cal Q}%
b=\omega \varepsilon ^\mu \partial _\mu \overline{c}+v^\mu \partial _\mu b\;.
\label{qanti}
\end{equation}
Thus, for the gauge fixing action we get\ 
\begin{eqnarray}
{{\cal S}_{gf}} &=&Q\int d^4x\;tr(\overline{c}\partial A)
\label{landau-g-fi} \\
&=&tr\displaystyle \int d^4x\;\left( b\partial A+\overline{c}\partial
Dc-\omega \overline{c}\partial \psi -\displaystyle \frac{\varepsilon ^\nu }2%
\overline{c}\partial ^\mu \chi _{\nu \mu }-\displaystyle \frac{\varepsilon
^\mu }8\overline{c}\partial _\mu \eta \right) \;,  \nonumber
\end{eqnarray}
so that the gauge fixed action $\left( {\cal S}_{TSYM}+S_{gf}\right) \;$is $%
{\cal Q}$-invariant. The above equation means that the gauge fixing
procedure has been worked out by taking into account not only the pure local
gauge symmetry but also the additional nonlinear invariances $\delta _{{\cal %
W}}\;$and $\delta _\mu $.

In order to obtain the Slavnov-Taylor identity\ we first couple the
nonlinear ${\cal Q}$-transformations of the fields $(c,\phi ,A,\psi ,%
\overline{\phi },\eta ,\chi )$ to a set of antifields $(c^{*},\phi
^{*},A^{*},\psi ^{*},\overline{\phi }^{*},\eta ^{*},\chi ^{*}),$

\begin{equation}
{\cal S}_{ext}=\displaystyle tr\int d^4x\;(\;\Phi ^{*i}{\cal Q}\Phi _i)\;,
\label{s-ext}
\end{equation}
where $\Phi ^i,$ $\Phi ^{*i}$ represent all fields and respective
antifields. Moreover, taking into account that the extended operator ${\cal Q%
}$ is nilpotent only modulo the equations of motion of the fields $\psi _\mu 
$ and $\;\chi _{\mu \nu }$, we also introduce a term quadratic in the
corresponding antifields $\psi ^{*\mu },\chi ^{*\mu \nu },\;${\it i.e.} 
\begin{equation}
{\cal S}_{quad}=tr\int d^4x\left( \frac{g^2}8\omega ^2\chi ^{*\mu \nu }\chi
_{\mu \nu }^{*}-\frac{g^2}4\omega \chi ^{*\mu \nu }\varepsilon _\mu \psi
_v^{*}-\frac{g^2}{32}\varepsilon ^\mu \varepsilon ^\nu \psi _\mu ^{*}\psi
_v^{*}+\frac{g^2}{32}\varepsilon ^2\psi ^{*\mu }\psi _\mu ^{*}\right) \;.
\label{s-quad}
\end{equation}
The complete action

\begin{equation}
\Sigma ={\cal S}_{TSYM}+{\cal S}_{gf}+{\cal S}_{ext}+{\cal S}_{quad}\;,
\label{c-action}
\end{equation}
obeys the classical Slavnov-Taylor identity 
\begin{equation}
\begin{array}{c}
{\cal S}(\Sigma )=\displaystyle tr\int d^4x\left( \frac{\delta \Sigma }{%
\delta \Phi ^{*i}}\frac{\delta \Sigma }{\delta \Phi _i}+(b+v^\mu \partial
_\mu \overline{c})\frac{\delta \Sigma }{\delta \overline{c}}\right. \\ 
\;\;\;\;\;\;\;\;\displaystyle \left. +(\omega \varepsilon ^\mu \partial _\mu 
\overline{c}+v^\mu \partial _\mu b)\frac{\delta \Sigma }{\delta b}\right)
-\omega \varepsilon ^\mu \displaystyle \frac{\partial \Sigma }{\partial
v^\mu }=0\;.
\end{array}
\label{tym-s-t-op}
\end{equation}
This equation will be the starting point for the analysis of the
renormalizability of the model.

At this point, it is worthwhile to draw the attention to a particular
feature of the complete action $\Sigma $ given by eq.$\left( \ref{c-action}%
\right) $. One should notice that, in this action, the global ghost
parameter $\omega $ only appears analytically. We expect therefore that the
physical sectors of the theory should be characterized by field polynomials
strictly analytic in $\omega $. This information will be of great relevance
when we come to the characterization of the possible counterterms and
anomalies of the theory.

The Slavnov-Taylor identity $\left( \ref{tym-s-t-op}\right) $ can be
simplified by using the fact that the complete action $\Sigma \;$is
invariant under space-time translations. Indeed, the dependence of $\Sigma
\; $on the corresponding translation constant ghost $v^\mu \;$turns out to
be fixed by the following linearly broken Ward identity 
\begin{equation}
\begin{array}{l}
\displaystyle \frac{\partial \Sigma }{\partial v^\mu }=\Delta _\mu ^{cl}=%
\displaystyle tr\int d^4x(c^{*}\partial _\mu c-\phi ^{*}\partial _\mu \phi
-A^{*\nu }\partial _\mu A_\nu +\psi ^{*\nu }\partial _\mu \psi _\nu -%
\overline{\phi }^{*}\partial _\mu \overline{\phi } \\ 
\;\;\;\;\;\;\;\;\;\; +\eta ^{*}\partial _\mu \eta +\displaystyle \frac
12\chi ^{*\nu \sigma }\partial _\mu \chi _{\nu \sigma })\;.
\end{array}
\label{v-break}
\end{equation}
This means that we can completely eliminate the global constant ghost $v^\mu 
$ without any further consequence. Introducing the action $\widehat{\Sigma }$
through

\begin{equation}
\Sigma =\widehat{\Sigma }+v^\mu \Delta _\mu ^{cl}\;,\;\;\;\;\;\;\;\frac{%
\partial \widehat{\Sigma }}{\partial v^\mu }=0\;,  \nonumber
\end{equation}
it is easily verified from $\left( \ref{tym-s-t-op}\right) $ that $\widehat{%
\Sigma }$ obeys the modified Slavnov-Taylor identity

\begin{equation}
{\cal S}(\widehat{\Sigma })\;{\cal =\;}\omega \varepsilon ^\mu \Delta _\mu
^{cl}\;\;.  \label{n-tym-s-t}
\end{equation}
Besides $\left( \ref{n-tym-s-t}\right) $, the classical action $\widehat{%
\Sigma }\;$turns out to be characterized by further additional constraints 
{\it \cite{book}}, namely$\;$the Landau gauge fixing condition, the
antighost equation, and the linearly broken ghost Ward identity (typical of
the Landau gauge), respectively 
\begin{equation}
\begin{array}{l}
\displaystyle \frac{\delta \widehat{\Sigma }}{\delta b}=\partial
A\;,\;\;\;\;\;\;\displaystyle \frac{\delta \widehat{\Sigma }}{\delta 
\overline{c}}+\partial _\mu \displaystyle \frac{\delta \widehat{\Sigma }}{%
\delta A_\mu ^{*}}=0\;, \\ 
\displaystyle \int d^4x\left( \frac{\delta \widehat{\Sigma }}{\delta c}%
+\left[ \overline{c},\frac{\delta \widehat{\Sigma }}{\delta b}\right]
\right) =\Delta _c^{cl}\;,
\end{array}
\label{idens}
\end{equation}
with $\Delta _c^{cl}\;$a linear classical breaking 
\begin{equation}
\Delta _c^{cl}=\int d^4x\;\left( [c,c^{*}]-[A,A^{*}]-[\phi ,\phi ^{*}]+[\psi
,\psi ^{*}]-[\overline{\phi },\overline{\phi }^{*}]+[\eta ,\eta ^{*}]+\frac
12[\chi ,\chi ^{*}]\right) \;.  \label{gh-lin-break}
\end{equation}
Following the standard procedure, let us introduce the so called reduced
action {\it \cite{book}} $\widetilde{{\cal S}}\;$ defined through the gauge
fixing condition $\left( \ref{idens}\right) $ as

\begin{equation}
\widehat{\Sigma }=\widetilde{{\cal S}}\;+\;tr\int d^4x\;b\partial A\;,
\label{red-act}
\end{equation}
so that $\widetilde{{\cal S}}$\ is independent from the Lagrangian
multiplier $b$. Moreover, from the antighost equation $\left( \ref{idens}%
\right) \;$it follows that $\widetilde{{\cal S}}\;$depends from the
antighost $\overline{c}\;$only through the combination$\;A_\mu ^{*}+\partial
_\mu \overline{c}\;$. From now on $\;A_\mu ^{*}$ will stand for this
combination. Accordingly, for the Slavnov-Taylor identity we get

\begin{eqnarray}
{\cal S}(\widetilde{{\cal S}})=tr\int d^4x\left( \frac{\delta \widetilde{%
{\cal S}}}{\delta \Phi ^{*i}}\frac{\delta \widetilde{{\cal S}}}{\delta \Phi
_i}\right) =\omega \varepsilon ^\mu \Delta _\mu ^{cl}\;\;.
\label{n-tym-st-op}
\end{eqnarray}
\ As a consequence the linearized Slavnov-Taylor operator ${\cal B}_{%
\widetilde{{\cal S}}}\;$defined as

\begin{eqnarray}
{\cal B}_{\widetilde{{\cal S}}}=tr\int d^4x\left( \frac{\delta \widetilde{%
{\cal S}}}{\delta \Phi ^i}\frac \delta {\delta \Phi _i^{*}}+\frac{\delta 
\widetilde{{\cal S}}}{\delta \Phi _i^{*}}\frac \delta {\delta \Phi
^i}\right) \;  \label{n-tym-lin-op}
\end{eqnarray}
is not strictly nilpotent. Instead, we have

\begin{equation}
{\cal B}_{\widetilde{{\cal S}}}{\cal B}_{\widetilde{{\cal S}}}=\omega
\varepsilon ^\mu {\cal P}_\mu \;,  \label{nil-lin-tym}
\end{equation}
meaning that ${\cal B}_{\widetilde{{\cal S}}}\;$is nilpotent only modulo a
total derivative. It follows then that ${\cal B}_{\widetilde{{\cal S}}}\;$%
becomes a nilpotent operator when acting on the space of the integrated
local polynomials in the fields and antifields. This is the case, for
instance, of the invariant counterterms and of the anomalies.

\section{Renormalization of the Twisted Theory}

We are now ready to discuss the renormalization of the twisted $N=2$
Yang-Mills theory. The first task is that of characterizing the cohomology
classes of the linearized Slavnov-Taylor operator which turn out to be
relevant for the anomalies and the invariant counterterms. Let us recall
that both anomalies and invariant counterterms are integrated local
polynomials $\Delta ^G\;$in the fields, antifields,$\;$and in the global
ghosts $(\omega ,\varepsilon )$,$\;$with dimension four, vanishing ${\cal R}$%
-charge and ghost number $G\;$respectively one and zero. In addition, they
are constrained by the consistency condition

\begin{equation}
{\cal B}_{\widetilde{{\cal S}}}\Delta ^G=0\;,\;\;\;\;\;\;\;\;\;\;\;G=0,1\;\;.
\label{int-cons-cond}
\end{equation}
In order to characterize the integrated cohomology of ${\cal B}_{\widetilde{%
{\cal S}}}$ we introduce the operator ${\cal N}_\varepsilon =\varepsilon
^\mu \partial /\partial \varepsilon ^\mu ,\;$which counts the number of
global ghosts $\varepsilon ^\mu \;$contained in a given field polynomial.
Accordingly, the functional operator ${\cal B}_{\widetilde{{\cal S}}}\;$%
displays the following $\varepsilon $-expansion

\begin{equation}
{\cal B}_{\widetilde{{\cal S}}}=b_{_{\widetilde{{\cal S}}}}+\varepsilon ^\mu 
{\cal W}_\mu +\frac 12\varepsilon ^\mu \varepsilon ^\nu {\cal W}_{\mu \nu
}\;,  \label{lin-dec}
\end{equation}
where, from eq.$\left( \ref{nil-lin-tym}\right) \;$the operators $b_{_{%
\widetilde{{\cal S}}}},\;{\cal W}_\mu ,\;{\cal W}_{\mu \nu }\;$are easily
seen to obey the following algebraic relations

\begin{equation}
b_{_{\widetilde{{\cal S}}}}b_{_{\widetilde{{\cal S}}}}=0\;,\;\;\;\;\;\left\{
b_{_{\widetilde{{\cal S}}}},{\cal W}_\mu \right\} =\omega {\cal P}_\mu \;,
\label{b-exact-nilp}
\end{equation}
\begin{equation}
\begin{array}[b]{l}
\left\{ {\cal W}_\mu ,{\cal W}_\nu \right\} +\left\{ b_{_{\widetilde{{\cal S}%
}}},{\cal W}_{\mu \nu }\right\} =0\;, \\ 
\left\{ {\cal W}_\mu ,{\cal W}_{\nu \rho }\right\} +\left\{ {\cal W}_\nu ,%
{\cal W}_{\rho \mu }\right\} +\left\{ {\cal W}_\rho ,{\cal W}_{\mu \nu
}\right\} =0\;, \\ 
\left\{ {\cal W}_{\mu \nu },{\cal W}_{\rho \sigma }\right\} +\left\{ {\cal W}%
_{\mu \rho },{\cal W}_{\nu \sigma }\right\} +\left\{ {\cal W}_{\mu \sigma },%
{\cal W}_{\nu \rho }\right\} =0\;.
\end{array}
\label{W-alg}
\end{equation}
>From $\left( \ref{b-exact-nilp}\right) \;$we observe that the operator $b_{_{%
\widetilde{{\cal S}}}}\;$is strictly nilpotent and that the vector operator $%
{\cal W}_\mu \;$allows to decompose the space-time translations ${\cal P}%
_\mu \;$as a $b_{_{\widetilde{{\cal S}}}}-$anticommutator, providing thus an
off-shell realization of the algebra $\left( \ref{wz-talg-1}\right) $. It is
just given by 
\begin{equation}
b_{_{\widetilde{{\cal S}}}}=s+\omega \delta _{{\cal W}}\;\;.  \label{bsw}
\end{equation}

According to the general results on cohomology, the integrated cohomology of
\ ${\cal B}_{\widetilde{{\cal S}}}\;$is isomorphic to a subspace of the
integrated cohomology of \ $b_{_{\widetilde{{\cal S}}}}$ {\it \cite{dixon}}.
Since \ $b_{_{\widetilde{{\cal S}}}}$ is exactly nilpotent, one can pass
from the integrated version of the Wess-Zumino consistency condition $\left( 
\ref{int-cons-cond}\right) $ to its local version, which leads to the
following set of descent equations,

\begin{eqnarray}
b_{_{\widetilde{{\cal S}}}}\Omega _4^G\;+\;\omega \partial ^\mu \Omega
_{\frac 72\mu }^G\; &=&\;0\;,  \label{desc-eqs} \\
b_{_{\widetilde{{\cal S}}}}\Omega _{\frac 72\mu }^G\;+\;\omega \partial ^\nu
\Omega _{3\left[ \mu \nu \right] }^G\; &=&\;0\;,  \nonumber \\
b_{_{\widetilde{{\cal S}}}}\Omega _{3[\mu \nu ]}^G\;+\;\omega \partial ^\rho
\Omega _{\frac 52[\mu \nu \rho ]}^G &=&\;0\;,  \nonumber \\
b_{_{\widetilde{{\cal S}}}}\Omega _{\frac 52[\mu \nu \rho ]}^G\;+\;\omega
\partial ^\sigma \Omega _{2[\mu \nu \rho \sigma ]}^G\; &=&\;0\;,  \nonumber
\\
b_{_{\widetilde{{\cal S}}}}\Omega _{2[\mu \nu \rho \sigma ]}^G\; &=&\;0\;, 
\nonumber
\end{eqnarray}
where the cocycle $\Omega _D^G$ has ghost number $G$ and dimension $D.$

The presence of the parameter $\omega $ in front of all the derivatives in
the above set of equations is a feature of the algebra given by $\left( \ref
{b-exact-nilp}\right) $. In fact, as we have commented before in the
preceding section, we are interested in the characterization of those
cohomologically nontrivial cocycles which are given by local field
polynomials depending analytically on the parameter $\omega $. In other
words, a cocycle will be nontrivial if it is analytic in $\omega $ and if it
cannot be written as a \ $b_{_{\widetilde{{\cal S}}}}$-variation of any
local field polynomial analytic in $\omega $. Now, using the algebra $\left( 
\ref{b-exact-nilp}\right) $, one can see that the solutions $\Omega _D^G$ of
the descent equations $\left( \ref{desc-eqs}\right) $ can be obtained by
suitably applying the operator ${\cal W}_\mu $ on the nontrivial solutions
of the local cohomology of $b_{_{\widetilde{{\cal S}}}}$ in each level of
the descent equations {\it \cite{st}}. Also, it is not difficult to show
that the operator ${\cal W}_\mu $ preserves the analyticity in $\omega $ of
the space where it acts upon, {\it i.e} it transforms local polynomials
analytic in $\omega $ into local polynomials analytic in $\omega $. Then, as
the nontrivial solutions of the local cohomology of $b_{_{\widetilde{{\cal S}%
}}}$ belong to this analytic space, it is assured that ${\cal W}_\mu $ will
map such solutions into nontrivial solutions of the cohomology of $b_{_{%
\widetilde{{\cal S}}}}$ modulo total derivatives. As a consequence, this
latter cohomology will also be restricted to the space of field polynomials
analytic in $\omega $.

We are interested in the solutions of the descent equations in the case of
the invariant counterterms and of the gauge anomalies, corresponding
respectively to the sectors of ghost number $G=0,1$. In the case of the
gauge anomalies, one can show that there is no possible nontrivial solution
for the local cohomology of $b_{_{\widetilde{{\cal S}}}}$ with the correct
quantum numbers at any level of the descent equations $\left( \ref{desc-eqs}%
\right) $. It is important to mention that this result, already obtained in 
{\it \cite{magg} } in the analysis of the $N=2$ untwisted gauge theories,
means that there is no possible extension of the nonabelian Adler-Bardeen
gauge anomaly compatible with $N=2$ supersymmetry.

In the case of the invariant counterterms, the analysis of the local
cohomology of $b_{_{\widetilde{{\cal S}}}}$ shows the existence of some
nontrivial solutions in different levels of the descent equations. The first
one has dimension $2$ and is given by 
\begin{equation}
\Delta \;=\frac 12tr\phi ^2\;.  \label{p2}
\end{equation}
It is a solution of the local cohomology of $b_{_{\widetilde{{\cal S}}}}$
for the last of the eqs.$\left( \ref{desc-eqs}\right) $. The second term has
dimension $3$, 
\begin{equation}
\Delta _{\mu \nu }=a\left( F_{\mu \nu }^{+}\phi +\frac{g^2\omega }2\chi
_{\mu \nu }^{*}\phi \right) \;,  \label{p3}
\end{equation}
and it is a solution of the local cohomology of $b_{_{\widetilde{{\cal S}}}}$
at the intermediate level of $\Omega _{3[\mu \nu ]}^0$. As we will discuss
in the following, this term is ruled out from the cohomology of the complete
operator ${\cal B}_{\widetilde{{\cal S}}}$ by requiring invariance under the
vector symmetry operator ${\cal W}_\mu $. Actually, it is possible to show
that the same happens for the other nontrivial solutions of the descent
equations $(\ref{desc-eqs})$, thus we do not report on them here.

Before analyzing in detail the consequences of the above results on the
cohomology of the complete operator ${\cal B}_{\widetilde{{\cal S}}}$, let
us discuss here the important issue of the analyticity in the constant
ghosts. In fact, the requirement of analyticity in the ghosts $(\varepsilon
_\mu ,\omega )$, stemming from pure perturbative considerations, is one of
the most important ingredients in the cohomological analysis that we are
doing. It is an almost trivial exercise to show that both cocycles $\left( 
\ref{p2}\right) $ and $\left( \ref{p3}\right) \;$can be expressed indeed as
a pure $b_{_{\widetilde{{\cal S}}}}-$variation, namely

\begin{equation}
\Delta =\frac 12b_{_{\widetilde{{\cal S}}}}tr\left( -\frac 1{\omega ^2}c\phi
+\frac 1{3\omega ^4}c^3\right) \;,  \label{triv-p2}
\end{equation}
\begin{equation}
\Delta _{\mu \nu }=ab_{_{\widetilde{{\cal S}}}}tr\left( \frac 1\omega \phi
\chi _{\mu \nu }\right) \;.  \label{triv-p3}
\end{equation}
These expressions illustrate in a very clear way the relevance of the
analyticity requirement. It is apparent from the eq.$\left( \ref{triv-p2}%
\right) \;$that the price to be payed in order to write $tr\phi ^2\;$as a
pure $b_{_{\widetilde{{\cal S}}}}-$variation is in fact the loss of
analyticity in the ghost $\omega $.

In other words, as long as one works in a functional space whose elements
are power series in the constant ghosts, the cohomology of $b_{_{\widetilde{%
{\cal S}}}}\;$is not empty. On the other hand, if the analyticity
requirement is relaxed, the cohomology of $b_{_{\widetilde{{\cal S}}}}$, and
therefore that of the complete operator ${\cal B}_{\widetilde{{\cal S}}}$,
becomes trivial, leading thus to the cohomological interpretation of
Baulieu-Singer {\it \cite{bs}} and Labastida-Pernici {\it \cite{lp}}. One
goes from the standard field theory point of view, of $N=2$ super
Yang-Mills, to the cohomological one, of the topological Yang-Mills theory,
by simply setting $\omega =1$, which of course implies that analyticity is
lost. In addition, it is rather simple to convince oneself that setting $%
\omega =1$ has the meaning of identifying the ${\cal R}$-charge with the
ghost number, so that the fields $(\chi ,\psi ,\eta ,\phi ,\overline{\phi }%
)\;$acquire a nonvanishing ghost number given respectively by $%
(-1,1,-1,2,-2) $. They correspond now to the so called topological ghosts of
the cohomological interpretation.

It is also interesting to point out that there is some relationship between
the analyticity in the global ghosts and the so called equivariant
cohomology proposed by {\it \cite{stora1,stora2}} in order to recover the
Witten's observables {\it \cite{dmpw,bc}}. Roughly speaking, the equivariant
cohomology can be defined as the restriction of the BRST\ cohomology to the
space of the gauge invariant polynomials which cannot be written as the BRST
variation of local quantities which are independent from the Faddeev-Popov
ghost $c$. Considering now the polynomial $tr\phi ^2$, we see that it yields
a nontrivial equivariant cocycle in the cohomological interpretation $%
(i.e.\;\omega =1)$, due to the unavoidable presence of the Faddeev-Popov
ghost $c$ on the right hand side of eq.$\left( \ref{triv-p2}\right) $.
However, the nontriviality of the second cocycle $\left( \ref{p3}\right) $
relies exclusively on the analyticity requirement.

In fact, it is not surprising at all to have found more than one independent
solution of the local cohomology of $b_{_{\widetilde{{\cal S}}}}$. One
should remember that the operator $\delta _{{\cal W}}\;$, which builds with $%
s$ the operator $b_{_{\widetilde{{\cal S}}}}$ of eq.$\left( \ref{bsw}\right) 
$ is not sufficient to completely fix the coefficients of the TSYM action $%
\left( \ref{tym}\right) $. One needs to impose the invariance under $\delta
_\mu $ in order to completely specify $\left( \ref{tym}\right) $. The
reflection of this point is the existence of others cocycles in the descent
equations for the operator $b_{_{\widetilde{{\cal S}}}}$. At this level,
this does not mean the existence of other $\beta $-functions, beyond that
associated to the gauge coupling $g$. Our interest, at the end, is in the
integrated cocycles invariant under ${\cal B}_{\widetilde{{\cal S}}}$ $%
\left( \ref{lin-dec}\right) $. Then, our approach is to climb the descent
equations $\left( \ref{desc-eqs}\right) $, which will give us the final
solution on the integrated cohomology of $b_{_{\widetilde{{\cal S}}}}$, and
afterwards demand invariance under ${\cal B}_{\widetilde{{\cal S}}}$ . By
using this last requirement we will get rid of the terms coming from the
cocycle $\left( \ref{p3}\right) $, as well as of that coming from the other
nontrivial cocycles.

In order to climb the descent equations, one can apply the operator ${\cal W}%
_\mu $, and reach the solution at the upper level for $\Omega _4^0$. This
solution, when integrated, can be written in the form 
\begin{eqnarray}
\Omega ^0 &=&\varepsilon ^{\mu \nu \rho \tau }{\cal W}_\mu {\cal W}_\nu 
{\cal W}_\rho {\cal W}_\tau \int d^4x\;\Delta \;+{\cal W}^\mu {\cal W}^\nu
\int d^4x\;\Delta _{\mu \nu }+b_{_{\widetilde{{\cal S}}}}-variation\;\; 
\nonumber \\
&=&{\cal S}_{TSYM}+a\Xi +b_{_{\widetilde{{\cal S}}}}\widetilde{\Omega }%
^{-1}\;,  \label{integ-coc}
\end{eqnarray}
where $\widetilde{\Omega }^{-1}$ is an arbitrary integrated polynomial in
the fields analytic in $\omega $ with ghost number $-1$, and

\begin{eqnarray}
\Xi &=&\int d^4x\left( \frac{g^2\omega }4F^{+\mu \nu }\chi _{\mu \nu }^{*}+%
\frac{g^4\omega ^2}8\chi _{\mu \nu }^{*}\chi ^{*\mu \nu }-\frac 14\chi ^{\mu
\nu }(D_\mu \psi _\nu -D_\nu \psi _\mu )^{+}\right.  \nonumber \\
&&\left. -\frac 14\phi \left\{ \chi ^{\mu \nu },\chi _{\mu \nu }\right\}
-\frac 34\;\psi ^\mu D_\mu \eta -\frac 34\omega g^2\phi D^\mu \psi _\mu
^{*}-\frac 34\omega g^2\psi ^\mu A_\mu ^{*}\right.  \nonumber \\
&&\left. +\frac 3{16}\phi \left\{ \eta ,\eta \right\} -\frac 32\omega
^2g^2\phi c^{*}+\frac 34\omega g^2\phi \left[ \overline{\phi },\eta
^{*}\right] \right) \;.  \label{int-sol}
\end{eqnarray}
Now, we need to impose the invariance of $\Omega ^0$ under ${\cal B}_{%
\widetilde{{\cal S}}}$ $\left( \ref{lin-dec}\right) $, which, in particular,
means invariance under ${\cal W}_\mu $ 
\begin{equation}
{\cal W}_\mu \Omega ^0={\cal W}_\mu \;\left( {\cal S}_{TSYM}+a\Xi +b_{_{%
\widetilde{{\cal S}}}}\widetilde{\Omega }^{-1}\right) =0\;.
\label{winvariance}
\end{equation}
Obviously, the action ${\cal S}_{TSYM}$ of eq.$\left( \ref{tym}\right) $ is
already invariant under ${\cal W}_\mu $. Then, using the algebra $\left( \ref
{W-alg}\right) $, the equation $\left( \ref{winvariance}\right) $ gives the
consistency condition 
\begin{equation}
a{\cal W}_\mu \Xi =b_{_{\widetilde{{\cal S}}}}\Lambda _\mu ^{-1}\;,
\label{wet}
\end{equation}
where $\Lambda _\mu ^{-1}$ is an arbitrary integrated polynomial in the
fields analytic in $\omega $ with ghost number $-1$. This implies that
either ${\cal W}_\mu \Xi $ is a trivial cocycle analytic in $\omega $, or
the coefficient $a$ has to vanish. One can show, by a straightforward
calculation, that the only way to write ${\cal W}_\mu \Xi $ as an exact
cocycle is to loose the analyticity in $\omega $

\begin{eqnarray*}
{\cal W}_\mu \Xi  &=&-\frac 1\omega b_{_{\widetilde{{\cal S}}}}\left( \frac
38F_{\mu \nu }^{-}D^\nu \overline{\phi }+\frac 3{64}\left[ \phi ,\overline{%
\phi }\right] D_\mu \overline{\phi }-\frac 38\psi ^\nu \left[ \chi _{\mu \nu
},\overline{\phi }\right] \right.  \\
&&\left. +\frac 3{32}\psi _\mu \left[ \eta ,\overline{\phi }\right] -\frac
1{16}\omega g^2\chi _{\mu \nu }^{*}D^\nu \overline{\phi }+\frac 14\omega
g^2\chi _{\mu \nu }A^{*\nu }\right.  \\
&&\left. -\frac 38\omega g^2F_{\mu \nu }^{-}\psi ^{*\nu }-\frac 3{64}\omega
g^2\psi _\mu ^{*}\left[ \phi ,\overline{\phi }\right] -\frac 34\omega
g^2\psi _\mu c^{*}+\frac 1{16}\omega ^2g^4\chi _{\mu \nu }^{*}\psi ^{*\nu
}\right) \;,
\end{eqnarray*}
{\it i.e.}, the only allowed solution for the equation $\left( \ref{wet}%
\right) $ is given by $a=0$. Then, going back to eq.$\left( \ref{integ-coc}%
\right) $, we can see that our cohomological analysis finally leads us to
the conclusion that the nontrivial part of the counterterm $\Omega ^0$ can
be written as the starting $N=2$ super-Yang-Mills action modulo a trivial $%
b_{_{\widetilde{{\cal S}}}}-term$%
\begin{equation}
\Omega ^0={\cal S}_{TSYM}+b_{_{\widetilde{{\cal S}}}}\widetilde{\Omega }%
^{-1}=\varepsilon ^{\mu \nu \rho \tau }{\cal W}_\mu {\cal W}_\nu {\cal W}%
_\rho {\cal W}_\tau \int d^4x\;\frac 12tr\phi ^2+b_{_{\widetilde{{\cal S}}%
}}-variation\;  \ ,\label{coc-final}
\end{equation}
namely, 
\begin{equation}
{\cal S}_{TSYM} \approx \varepsilon ^{\mu \nu \rho \tau }{\cal W}_\mu {\cal W}_\nu {\cal W}%
_\rho {\cal W}_\tau \int d^4x\;\frac 12tr\phi ^2 \ ,  \label{cocfinal}
\end{equation}
up to trivial cocycles.

The fact that we were left with only one arbitrary coefficient in the
nontrivial part of the counterterm (which is the global coefficient of $%
{\cal S}_{TSYM}$) means the presence of only one coupling in the theory, and
consequently, of only one $\beta $-function for the twisted $N=2$ Yang-Mills
theory.

\section{Conclusion}

We have shown how the quantization of the twisted $N=2$ super-Yang-Mills
action can be done by taking into account the full $N=2\;$twisted$\;$%
supersymmetric algebra. Nontrivial cohomology classes are characterized by
requiring analyticity in the twisted constant global ghosts of the $N=2$
quantized theory. The analyticity requirement, following from perturbation
theory, plays a crucial r\^{o}le as it defines a criterium in order to have
a nontrivial cohomology.

We have also seen that the operator $b_{_{\widetilde{{\cal S}}}}$ has a
nonvanishing integrated analytic cohomology only in the sector of the
invariant counterterms, coming from the nontrivial local elements given by $%
\left( \ref{p2}\right) $ and $\left( \ref{p3}\right) $. Finally, we have
been able to show that the cohomology of ${\cal B}_{\widetilde{{\cal S}}}$
in the sector of the invariant counterterms contains a unique element given
in eq.$\left( \ref{cocfinal}\right) $. This result is in complete agreement
with that found in the case of untwisted $N=2$ YM {\it \cite{magg}}.

Moreover, as pointed out by eq.$\left( \ref{cocfinal}\right) $, we emphasize
that the origin of the $N=2$ super-Yang-Mills action $\left( \ref{tym}%
\right) $ can be traced back, modulo an irrelevant exact cocycle, to the
invariant polynomial $tr\phi ^2$, eq.$\left( \ref{p2}\right) $. Let us also
recall here that the explicit Feynman diagrams computation yields a
nonvanishing value for the renormalization of the gauge coupling, meaning
that the twisted{\it \ }version of{\it \ }$N=2${\it \ }Yang-Mills{\it \ }%
possesses a nonvanishing $\beta $-function. Of course, the latter agrees
with that of the pure $N=2$ untwisted Yang-Mills {\it \cite{beta}}.
Moreover, it is well known that the $\beta $-function of $N=2$\thinspace
Yang-Mills theory receives only one loop order contributions {\it \cite
{beta1}}. On the other hand it is known since several years that in the $N=2$
susy gauge theories the Green's functions with the insertion of composite
operators of the kind of the invariant polynomials of the form $tr\phi ^n$
display remarkable finiteness properties and can be computed exactly, even
when nonperturbative effects are taken into account {\it \cite{akmrv}}. It
is natural therefore to guess that the finiteness properties of $tr\phi
^2$ are at the origin of the absence of the higher order corrections for the
gauge $\beta $-function of both twisted and untwisted $N=2$ gauge theories.

Finally, it is worth mentioning that, recently, an algebraic proof of the nonrenormalization theorem of the $N=2\;$ $\beta $-function has been achieved {\it \cite{alb}} along the lines of the present work. As expected, the relationship $\left( \ref{cocfinal}\right)$ between the $N=2$ Yang-Mills action and the local gauge invariant operator $tr\phi^2$ turns out to play a pivotal role.

\vspace{5mm}

{\Large {\bf Acknowledgements}} The financial supports from CNPq - Brazil,
SR-2 UERJ and from FAPERJ-Rio are gratefully acknowledged.


\begin{thebibliography}{99}
\bibitem{tym}  E. Witten, {\bf Comm. Math. Phys. 117 (1988) 353;}

\bibitem{mar}  M. Mari\~{n}o, ``{\it The Geometry of Supersymmetric Gauge
Theories in Four Dimensions'', }{\bf US-Ft-2/97, hep-th/9701128;}

\bibitem{lp}  J.M.F. Labastida and M. Pernici, {\bf Phys. Lett. B212 (1988)
56;}

\bibitem{bs}  L. Baulieu and I.M. Singer, {\bf Nucl. Phys. B15 (1988) 12;}\\%
L. Baulieu and I.M. Singer,{\bf \ Comm. Math. Phys. 135 (1991) 253;}

\bibitem{stora1}  S. Ouvry, R. Stora and P. Van Baal, {\bf Phys. Lett. B220
(1989) 159;}

\bibitem{stora2}  J. Kalkman, {\bf Comm. Math. Phys. 153 (1993) 447;}\\R.
Stora, ``{\it Equivariant Cohomology and Topological Theories'', }in BRS
Symmetry, M. Abe, N. Nakanishi, Iojima eds., {\bf Universal Academy Press,
Tokyo, Japan, 1996;}\\R. Stora, F. Thuillier and J.C. Wallet, ``{\it %
Algebraic Structure of Cohomological Field Theory Models and Equivariant
Cohomology'',} lectures given at the First Caribbean Spring School of
Mathematics and Theoretical Physics, R. Coquereaux, M. Dubois-Violette and
P. Flad Eds., {\bf World Scientific Publ., 1995;}\\R. Stora,{\it \
``Exercises in Equivariant Cohomology}''{\it ,}{\bf \ ENSLAPP-A-619/96,
hep-th/9611114};\\R. Stora, ``{\it De la fixation de jauge consideree comme
un des beaux arts et de la symetrie de Slavnov qui s 'ensuit'', }{\bf %
ENSLAPP-A-620/96, hep-th/9611115;}\\R. Stora,{\it \ ``Exercises in
Equivariant Cohomology and Topological Theories} ''{\it ,}{\bf \
hep-th/9611116;}

\bibitem{white}  P. L. White, {\bf Class. Quantum Grav. 9 (1992) 413;}\\P.
L. White, {\bf Class. Quantum Grav. 9 (1992) 1663;}

\bibitem{magg}  N. Maggiore, {\bf Int. J. Mod. Phys. A10 (1995) 3781;}\\N.
Maggiore, {\bf Int. J. Mod. Phys. A10 (1995) 3937;}

\bibitem{horne} J. Horne, {\bf Nucl. Phys. B318 (1989) 22;} 

\bibitem{al} M. Alvarez and J.M.F. Labastida, {\bf Nucl. Phys. B437 (1995) 356;}

\bibitem{beta}  R. Brooks, D. Montano and J. Sonnenschein, {\bf Phys. Lett.
B214 (1988) 91;}

\bibitem{tft}  F. Delduc, C. Lucchesi, O. Piguet and S.P. Sorella, {\bf %
Nucl. Phys. B346 (1990) 313;}\\E. Guadagnini, N. Maggiore and S.P. Sorella, 
{\bf Phys. Lett. B255 (1991) 65;}\\N. Maggiore and S.P. Sorella, {\bf Int.
Journ. Mod. Phys. A8 (1993) 929;}\\C. Lucchesi, O. Piguet and S.P. Sorella, 
{\bf Nucl. Phys. B395 (1993) 325;}\\L.C.Q. Vilar, O. S. Ventura, C.A.G.
Sasaki and S.P. Sorella, ``{\it Algebraic Characterization of Vector
Supersymmetry in Topological Field Theories'',}{\bf \ CBPF-NF-007/97, to
appear in Journ. Math. Phys.;}

\bibitem{brt}  D. Birmingham, M. Blau, M. Rakowski and G. Thompson, {\bf %
Phys. Rep. 209 (1991) 129;}

\bibitem{vitoria}  F. Fucito, A. Tanzini, L.C.Q. Vilar, O.S. Ventura, C.A.G.
Sasaki and S.P. Sorella, ``{\it Perturbative Twisted Considerations on
Topological Yang-Mills Theory and on N=2 Supersymmetric Gauge Theories'', }%
lectures given at the First School on Field Theory and Gravitation of
Vitoria, {\bf CBPF-NF-043/97, hep-th/9707209;}

\bibitem{book}  O. Piguet and S.P. Sorella, {\it Algebraic Renormalization,} 
{\bf Monographs Series, m 28, Springer Verlag, Berlin, 1995;}

\bibitem{dixon}  J. A. Dixon, {\bf Comm. Math. Phys. 139 (1991) 495;}

\bibitem{st}  S.P. Sorella, {\bf Comm. Math. Phys. 157 (1993) 231;}\\S.P.
Sorella and L. Tataru, {\bf Phys. Lett. B324 (1994) 351;}

\bibitem{dmpw}  F. Delduc, N. Maggiore, O. Piguet and S. Wolf, {\bf Phys.
Lett. B385 (96) 132;}

\bibitem{bc}  A. Blasi and R. Collina, {\bf Phys. Lett. B222 (1989) 159;}

\bibitem{beta1}  M. Sohnius and P. West, {\bf Phys. Lett. B100 (1981) 45;\\}%
P. Howe, K. Stelle and P. West, {\bf Phys. Lett. B124 (1983) 55;}

\bibitem{akmrv}  D. Amati, K. Konishi, Y. Meurice, G.C. Rossi and G.
Veneziano, {\bf Phys. Rep. 162 (4) (88) 169.}

\bibitem{alb} A. Blasi, V.E.R. Lemes, N. Maggiore, S.P. Sorella, A. Tanzini, O.S. Ventura, L.C.Q. Vilar, {\it Perturbative Beta Function of N=2 Super Yang-Mills Theories}, {\bf hep-th/0004048}. 
\end{thebibliography}
\end{document}